\documentclass[pra,twoside,showpacs,twocolumn]{revtex4}
\usepackage{times}
\usepackage{amssymb}
\usepackage{amsmath}
\usepackage{graphicx}
\usepackage{epsfig}
\usepackage{dcolumn}
\usepackage{color}
\usepackage{bm}

\begin{document}

\title{External field effect on quantum features of radiation emitted by a
QW in a microcavity}
\author{Eyob A. Sete}
\email{eyobas@physics.tamu.edu}
\author{ Sumanta Das}
\affiliation{Institute for Quantum Science and Engineering and
Department of Physics and Astronomy, Texas A$\&$M University,
College Station, TX 77843-4242, USA}
\author{H. Eleuch}
\affiliation{Department of Physics and Astronomy, College of
Science, P. O. Box 2455, King Saud University, Riyadh 11451, Saudi
Arabia }
\date{\today}

\begin{abstract}
We consider a semiconductor quantum well in a microcavity driven by
coherent light and coupled to a squeezed vacuum reservoir. By
systematically solving the pertinent quantum Langevin equations in
the strong coupling and low excitation regimes, we study the effect
of exciton-photon detuning, external coherent light, and the
squeezed vacuum reservoir on vacuum Rabi splitting and on quantum
statistical properties of the light emitted by the quantum well. We
show that the exciton-photon detuning leads to a shift in polariton
resonance frequencies and a decrease in fluorescence intensity. We
also show that the fluorescent light exhibits quadrature squeezing
which predominately depends on the exciton-photon detuning and the
degree of the squeezing of the input field.
\end{abstract}
\pacs{42.55.Sa, 78.67.De, 42.50.Dv, 42.50.Lc} \maketitle

\section{Introduction}
Study of radiation-matter interaction, in two level quantum
mechanical systems have lead to several fascinating phenomena like
the Autler-Townes doublet \cite{Aut55}, vacuum Rabi splitting
\cite{Aga84,Tho92, Khi06}, antibunching and squeezing
\cite{Gar86,Car87,Vya92}.  In particular, interaction of two-level
atoms in a cavity with a coherent source of light and coupled to a
squeezed vacuum has been extensively studied
\cite{Aga89,Par90,Cir91,Rice96,Ge095,Smy96,Cle20,Str05,Set08}.
Currently there is renewed interest in such studies from the context
of semiconductor systems like quantum dots (QDs) and wells (QWs)
\cite{Baa04,Qua05,Hic108,Hic208, Set10} given their potential
application in opto-electronic devices \cite{Shi07}. In this regard,
intersubband excitonic transitions which have similarities to two
level atomic system has been primarily exploited.  However, it is
important to understand that the quantum nature of fluorescent light
emitted by excitons in QWs embedded inside a microcavity somewhat
differs to that of atomic cavity QED predictions. For example,
unlike antibunching observed in atoms embedded in a cavity a QW
exhibits bunching effects in the fluorescent spectrum of the emitted
radiation \cite{Vya20,Ere03}.  Further in the strong coupling
regime, for a resonant microcavity-QW interaction, exciton-photon
mode splitting and oscillatory excitonic emission has been
demonstrated \cite{Wei92,Pau95,Jac95}. Recently effect of a
non-resonant strong drive on the Intersubband excitonic transition
has been investigated and observation of Autler Townes doublets were
reported \cite{Dyn05,Car05,Wag2010}. In light of these new results
an eminent question of interest is thus, how does the quantum
nature of radiation emitted from a QW in a microcavity get effected
in presence of a squeezed vacuum environment and non-resonant drive?
We investigate this in the current paper.

We explore the interaction between an external field and a QW placed
in a microcavity coupled to a squeezed vacuum reservoir. Our
analysis is restricted to the weak excitation regime where the
density of excitons is small. This allows us to neglect any
exciton-exciton interaction thereby simplifying our problem
considerably and yet preserving the physical insight. We further
assume the cavity-exciton interaction to be strong, which brings in
interesting features. Note that we consider the external field to be
in resonance with the cavity mode throughout the paper. We analyze
the effect of exciton-photon detuning, external coherent light, and
the squeezed reservoir on the quantum statistical properties and
polariton resonances in the strong coupling and low excitation
regimes. The effect of the coherent light on the behavior of the dynamical
evolution of the intensity fluorescent light is remarkably different to that of the
squeezed vacuum reservoir due to the nature of photons generated by the two systems.
This is due to the distinct nature of the photons
generated by the coherent and squeezed inputs. This effect is
manifested on the intensity of the fluorescent light. Both sources
lead to excitation of two or more excitons in the quantum well
creating a probability for emission of two or more photons
simultaneously. As a result of this, the photons tend to propagate
in bunches other than at random. Moreover, the fluorescent light
emitted by exciton in the quantum well exhibits nonclassical
property namely, quadrature squeezing.

\section{Model and Quantum Langevin equations}
\begin{figure}[t]
\includegraphics[width=8cm]{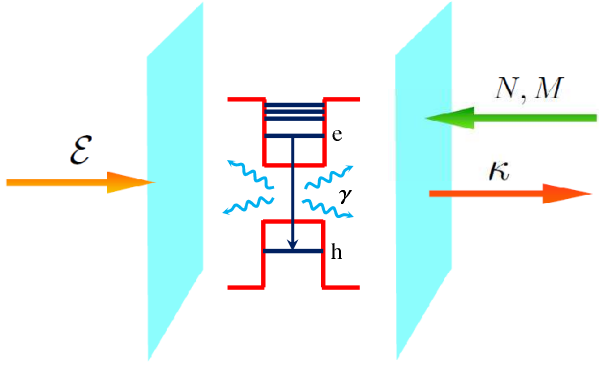}
\caption{Schematic representation of a quantum well (QW) in a driven
cavity coupled to a squeezed vacuum reservoir.} \label{fig0}
\end{figure}

We consider a semiconductor quantum well (QW) in a cavity driven by
external coherent light and coupled to a single mode squeezed vacuum
reservoir. The scheme is outlined in Fig. \ref{fig0}. In this work,
we are restricted to a linear regime in which the density of
excitons is small so that exciton-exciton scattering can be ignored.
We assume that the driving laser is at resonance with the cavity
mode while the exciton transition frequency is off resonant with the
cavity mode by an amount $\Delta=\omega_{0}-\omega_{c}$ with
$\omega_{0}$ and $\omega_{c}$ being the exciton and cavity mode
frequencies. The interaction of the cavity mode with the resonant
pump field and the exciton is described, in the rotating wave and
dipole approximations, by the Hamiltonian
\begin{equation}
H=\Delta b^{\dagger}b+i\varepsilon
(a^{\dagger}-a)+i\text{g}(a^{\dagger}b-ab^{\dagger})+H_{\text{loss}},
\end{equation}
where $a$ and $b$, are the annihilation operators for the cavity and
exciton modes satisfying the commutation relation
$[a,a^{\dagger}]=[b,b^{\dagger}]=1$ respectively; $\text{g}$ is the
photon-exciton coupling constant; $\varepsilon $, assumed to be real
and constant, is proportional to the amplitude of the pump field,
and $H_{\text{loss}}$ is the Hamiltonian that describes the
interaction of the exciton with the vacuum reservoir and also the
interaction of the cavity mode with the squeezed vacuum reservoir.

The quantum Langevin equations, taking into account the dissipation
processes, can be written as
\begin{equation}\label{2}
\frac{da}{dt}=-\frac{\kappa}{2} a+\text{g}b+\varepsilon+F(t),
\end{equation}
\begin{equation}\label{3}
\frac{db}{dt}=-(\frac{\gamma}{2}+i\Delta)b-\text{g}a+G(t),
\end{equation}
where $\kappa$ and $\gamma$ are cavity mode decay rate via the port
mirror and spontaneous emission decay rate for the exciton,
respectively;  $F=\sqrt{\kappa}F_{\text{in}}$ and
$G=\sqrt{\gamma}G_{\text{in}}$ with $F_{\text{in}}$ and
$G_{\text{in}}$ being the Langevin noise operators for the cavity
and exciton modes, respectively. Both noise operators have zero
mean, i.e., $\langle F_{\text{in}}\rangle=\langle
G_{\text{in}}\rangle=0$. For a cavity mode coupled to a squeezed
vacuum reservoir, the noise operator $F_{\text{in}}(t)$ satisfies
the following correlations:
\begin{equation}\label{4}
\left\langle F_{\text{in}}(t)F_{\text{in}}^{\dagger}(t^{\prime
})\right\rangle = (N+1) \delta (t-t^{\prime }),
\end{equation}
\begin{equation}\label{5}
\left\langle F_{\text{in}}^{\dagger}(t)F_{\text{in}}(t^{\prime
})\right\rangle =N\delta (t-t^{\prime }),
\end{equation}
\begin{equation}
\left\langle F_{\text{in}}(t)F_{\text{in}}(t^{\prime })\right\rangle
=\left\langle
F_{\text{in}}^{\dagger}(t)F_{\text{in}}^{\dagger}(t^{\prime
})\right\rangle= M\delta (t-t^{\prime }),
\end{equation}
where $N=\sinh^2r$ and $M=\sinh r\cosh r$ with $r$ being the squeeze
parameter characterize the mean photon number and the phase
correlations of the squeezed vacuum reservoir, respectively.
Further, the exciton noise operator $G_{\text{in}}$ satisfies the
following correlations:
\begin{equation}\label{7}
\left\langle G_{\text{in}}(t)G_{\text{in}}^{\dagger}(t^{\prime
})\right\rangle =\delta (t-t^{\prime }),
\end{equation}
\begin{equation}\label{8}
\left\langle G_{\text{in}}^{\dagger}(t)G_{\text{in}}(t^{\prime
})\right\rangle =\left\langle
G_{\text{in}}(t)G_{\text{in}}(t^{\prime })\right\rangle
=\left\langle
G_{\text{in}}^{\dagger}(t)G_{\text{in}}^{\dagger}(t^{\prime
})\right\rangle=0.
\end{equation}

Following the method outlined in \cite{Set10}, we obtain the
solution of the quantum Langevin equations \eqref{2} and \eqref{3}
in the strong coupling regime ($g\gg \gamma, \kappa$) to be
\begin{align}\label{a1}
    a(t)&=
    \eta_{1}(t)\varepsilon+\eta_{+}(t)a(0)+\eta_{3}(t)b(0)\notag\\
   &+\int_{0}^{t}dt^{\prime}\eta_{+}(t-t^{\prime})F(t^{\prime})+\int_{0}^{t}dt^{\prime}\eta_{3}(t-t^{\prime})G(t^{\prime}),
\end{align}
\begin{align}\label{a2}
    b(t)&=\eta_{-}(t)b(0)
    -\eta_{4}(t)\varepsilon-\eta_{3}(t)a(0)\notag\\
    &-\int_{0}^{t}dt^{\prime}\eta_{3}(t-t^{\prime})F(t^{\prime})+\int_{0}^{t}dt^{\prime}\eta_{-}(t-t^{\prime})G(t^{\prime}),
\end{align}

where
\begin{align}\label{a3}
    \eta_{1}(t)=\frac{1}{\mu}\sin \mu t~
    e^{-(\Gamma+i\Delta/2)t},
\end{align}
\begin{align}\label{a4}
    \eta_{\pm}(t)=\frac{1}{2\mu}\left[2\mu\cos \mu t\pm i\Delta \sin \mu
t\right]e^{-(\Gamma+i\Delta/2)t},
\end{align}
\begin{equation}\label{a5}
   \eta_{3}(t)=\frac{g}{\mu}\sin \mu t
    e^{-(\Gamma+i\Delta/2)t},
\end{equation}
\begin{align}\label{a7}
   \eta_{4}(t)=\frac{1}{g}-\frac{1}{g}\left[\cos (\mu t)+\frac{i\Delta}{2\mu}\sin\mu
   t)\right]
  e^{-(\Gamma+i\Delta/2)t},
\end{align}
where $\Gamma=(\kappa+\gamma)/4$ and $\mu=\sqrt{g^2+\Delta^2/4}$.
Using these solutions we study the
dynamical behavior of intensity, power spectrum, second-order
correlation function, and quadrature variance for the fluorescent
light emitted by the quantum well in the following sections.

\section{Intensity of fluorescent light}
In this section we analyze the properties of the fluorescent light
emitted by excitons  in the quantum well. In particular, we study
the effect of the external driving field, exciton-photon detuning,
and the squeezed vacuum reservoir. Note that the intensity of the
fluorescent light is proportional to the mean number of excitons. To
this end, the intensity of the fluorescent light can be expressed in
terms of the solutions of the Langevin equations as
\begin{equation}\label{m1}
    \langle
    b^{\dagger}b\rangle=\varepsilon^2|\eta_{4}(t)|^2+|\eta_{-}(t)|^2+\kappa
    N\int_{0}^{t}|\eta_{3}(t-t')|^2dt'.
\end{equation}
Here we have assumed the cavity mode is initially in vacuum state
[$\langle a^{\dagger}(0)a(0)\rangle=0$] while the quantum well
initially contains only one exciton [$\langle
b^{\dagger}(0)b(0)\rangle=1$]. In \eqref{m1} the first term
corresponds to the contribution from the external coherent light
while the last term is due to the the squeezed vacuum reservoir. It
is also easy to see that the intensity is independent of the
parameter  $M$, which characterizes the phase correlations of the
reservoir, implying that the same result could be obtained if the
cavity mode is coupled to a thermal reservoir.

Performing the integration using Eqs. \eqref{a5}-\eqref{a7}, we
obtain
\begin{align}\label{m2}
    \langle
    b^{\dagger}b\rangle&=\frac{\varepsilon^2}{g^2}+\frac{g^2\kappa
    N}{4\Gamma\mu^2}+(\lambda_{1}+\kappa N\lambda_{2})e^{-2\Gamma t}\notag\\
    &+\frac{\varepsilon^2}{g^2}\left(\lambda_{1}e^{-2\Gamma
t}-2\lambda_{3}e^{-\Gamma t}\right),
\end{align}
where
\begin{equation}\label{m3}
 \lambda_{1}(t)=\frac{\Delta^2}{4\mu^2}\sin^2\mu t+\cos^2\mu t,
   \end{equation}
\begin{align}\label{m4}
 \lambda_{2}(t)=-\frac{g^2
    }{4\mu^2}\Big(\frac{1}{\Gamma}+\frac{1}{\mu}\sin 2\mu
    t\Big),
\end{align}
\begin{equation}\label{m5}
 \lambda_{3}(t)=\cos\mu t\cos(\Delta t/2)+\frac{\Delta}{2\mu}\sin\mu t\sin(\Delta t/2).
\end{equation}
We immediately see that the intensity of the fluorescent light
reduces in the steady state to
\begin{align}\label{m2}
    \langle
    b^{\dagger}b\rangle_{ss}=\frac{\varepsilon^2}{g^2}+\frac{g^2\kappa
    N}{4\Gamma\mu^2}.
\end{align}
As expected the steady state intensity is inversely proportional to
the decay rate. The higher the decay rate the lower the intensity
and vise versa.
\begin{figure}[t]
\includegraphics{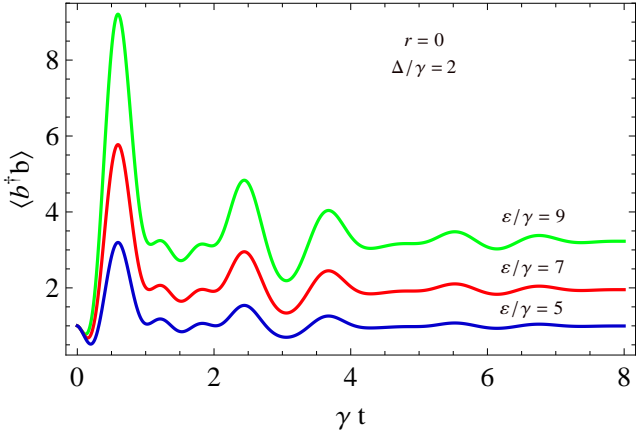}
\caption{Plots of the fluorescent intensity [Eq. \eqref{ee}] vs
scaled time $\gamma t$ for $\kappa/\gamma=1.2$, $g/\gamma=5$,
$\Delta/\gamma=2$, and for different values of
$\varepsilon/\gamma$.} \label{fig1}
\end{figure}
\begin{figure}[h]
\includegraphics{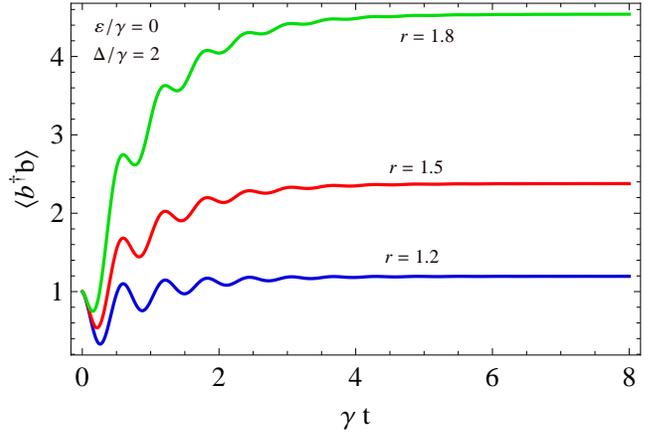}
\caption{Plots of the fluorescent intensity [Eq. \eqref{nn}] vs
scaled time $\gamma t$ for $\kappa/\gamma=1.2$, $g/\gamma=5$,
$\Delta/\gamma=2$, not external driving field
($\varepsilon/\gamma=0$) and for different values of $r$.}
\label{fig2}
\end{figure}

In order to clearly see the effect of the external coherent light on
the intensity, we set $N=0$ in Eq. \eqref{m2} and obtain
\begin{equation}\label{ee}
\langle
b^{\dagger}b\rangle=\frac{\varepsilon^2}{g^2}(1+\lambda_{1}e^{-2\Gamma
t}-2\lambda_{3}e^{-\Gamma t}).
\end{equation}
Figure \ref{fig1} shows the dependence of the intensity of the
fluorescent light on the external coherent field. In general, the
intensity increases with the amplitude of the pump field and
exhibits non periodic damped oscillations. Although there is a
decrease in the mean number of excitons for the initial moment,
cavity photons gradually excite one or more excitons in the quantum
well leading to enhanced emission of fluorescence. However, the
excitation of excitons saturates as time progresses limited by the
strength of the applied field. From the steady state intensity,
$\varepsilon^2/g^2$, we easily see that the field strength has to
exceed the exciton-photon coupling constant in order to see more
than one exciton in the quantum well in the long time limit.

On the other hand, the effect of the squeezed vacuum can be studied
by turning off the external driving field. Thus setting
$\varepsilon=0$ in Eq. \eqref{m2}, we get
\begin{equation}\label{nn}
\langle b^{\dagger}b\rangle=\frac{g^2\kappa
    N}{4\Gamma\mu^2}+(\lambda_{1}+\kappa N\lambda_{2})e^{-2\Gamma t}.
\end{equation}
In Fig. \ref{fig2}, we plot the intensity of the light emitted by
the exciton [Eq. \eqref{nn}] as a function of scaled time $\gamma t$
for a given photon-exciton detuning. This figure illustrates the
dependence of the intensity on squeezed vacuum reservoir impinging via the
partially transmitting mirror. Here also the intensity exhibits
damped oscillations at frequency $2\mu=2(g^2+\Delta^2/4)^{1/2}$
indicating exchange of energy between the excitons and cavity mode.
The intensity decreases at the initial moment and gradually
increases to steady state values $g^2\kappa N/4\Gamma \mu^2$.
While the intensity increases it shows oscillatory behavior which ultimately disappears for longer times.
Unsurprisingly, the intensity increases with the number of photons
coming in through the mirror. Comparing Figs. \ref{fig1} and
\ref{fig2}, we note that the intensity has different behavior for
the two cases. This can be explained in terms of the nature of
photon each source is producing. In the case of coherent light, the
photon distribution is Poisson and the photons propagate randomly.
This leads to uneven excitation of excitons that results in
nonperiodic oscillatory nature of the intensity. In the case of
squeezed vacuum source, however, the photons show bunching property
and hence can excite two or more excitons at the same time. This in
turn implies that depending on the strength of the impinging
squeezed vacuum field, there will be one or more
excitons in the quantum well.

\begin{figure}[t]
\includegraphics{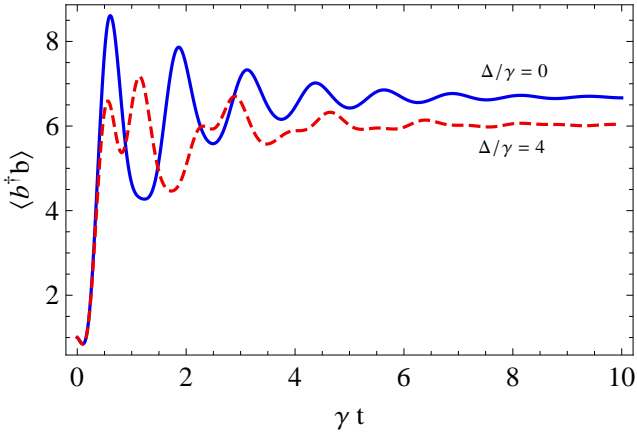}
\caption{Plots of the fluorescent intensity [Eq. \eqref{m2}] vs
scaled time $\gamma t$ for $\kappa/\gamma=1.2$, $g/\gamma=5$,
$r=1.8$, pump amplitude $\varepsilon/\gamma=7$ and for different
values of exciton-photon detuning ($\Delta/\gamma$).} \label{fig3}
\end{figure}

For the sake of completeness we further consider the effect of
detuning on the intensity of the fluorescent light at a given pump
field strength and squeezed photons. Figure \ref{fig3} shows the
intensity as a function of scaled time $\gamma t$. When the photon
is out of resonance with the exciton frequency there will be less
number of excitons in the quantum well and hence the fluorescent
intensity decreases. This is clearly shown in Fig. \ref{fig3}.

\section{Power spectrum}
The power spectrum of the fluorescent light in the steady state is
given by
\begin{equation}\label{s1}
    S(\omega)=\frac{1}{\pi}\text{Re}\int_{0}^{\infty}d\tau~e^{i(\omega-\omega_{0})
    \tau}\langle b^{\dagger}(t)b(t+\tau)\rangle_{ss},
\end{equation}
where $ss$ stands for steady state. The two time correlation
function that appears in the above integrand is found to be
\begin{align}\label{s2}
    \langle
    b^{\dagger}(t)b(t+\tau)\rangle_{ss}&=\frac{\varepsilon^2}{g^2}+\frac{\kappa N
    g^2}{4\Gamma\mu^3}e^{-(\Gamma+2i\Delta/2)\tau}\notag\\
    &\times(\mu\cos
    \mu\tau+\Gamma\sin\mu\tau).
\end{align}
Now employing this result in \eqref{s1} and performing the resulting
integration and carrying out the straightforward arithmetic, we
obtain
\begin{equation}\label{s3}
    S(\omega)=\frac{\varepsilon^2}{2\pi
    g^2}\delta(\omega-\omega_{0})+S_{\text{incoh}}(\omega),
\end{equation}
where
\begin{align}\label{s4}
    S_{\text{incoh}}(\omega)&=\frac{kNg^2}{16\pi\mu^3}\Big\{\frac{\Delta+4\mu-2\omega}{\Gamma^2+[\frac{\Delta}{2}+\mu-(\omega-\omega_{0})]^2}\notag\\
    &
      +\frac{-\Delta+4\mu+2\omega}{\Gamma^2+[\frac{\Delta}{2}-\mu-(\omega-\omega_{0})]^2}\Big\}.
\end{align}

We note that the power spectrum
has two components: coherent and incoherent parts. The coherent
component is represented by the delta function, which indeed
corresponds to the coherent light. The incoherent component given by
\eqref{s4}, arises as a result of the squeezed photons coming
through the port mirror. From Eq. \eqref{s4} it is clear that the
spectrum of the incoherent light is composed of two Lorentzians
having the same width $\Gamma$ but centered at two different
frequencies:$\omega-\omega_{0}=\mu+\Delta/2$ and
$\omega-\omega_{0}=\mu-\Delta/2 $. We then see that the detuning
leads to a shift in the resonance frequency components observed in
zero detuning ($\Delta=0)$.
\begin{figure}[t]
\includegraphics{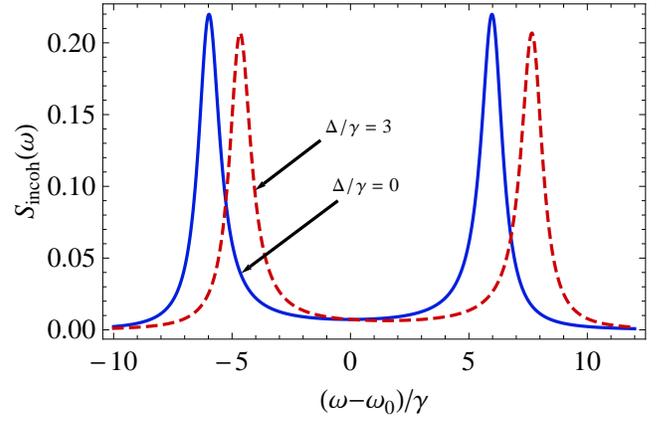}
\caption{Plots of the incoherent component of the power spectrum
[Eq. \eqref{s4}] vs scaled frequency $(\omega-\omega_{0})/\gamma$
for $\kappa/\gamma=1.2$, $g/\gamma=6$, squeeze parameter $r=1$, and
for different values of detuning, $\Delta/\gamma$.} \label{fig4}
\end{figure}
\begin{figure}[t]
\includegraphics[width=6cm]{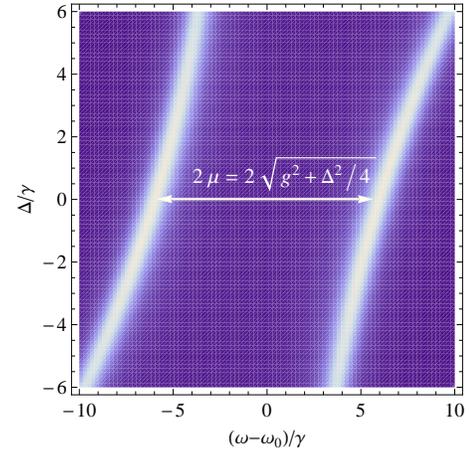}
\caption{Density plot of the incoherent component of the power
spectrum [Eq. \eqref{s4}] vs scaled frequency
$(\omega-\omega_{0})/\gamma$ and $\Delta/\gamma$ for
$\kappa/\gamma=1.2$, $g/\gamma=6$, and for squeeze parameter $r=1$.}
\label{fig5}
\end{figure}

\begin{table*}\label{tab1}
\begin{ruledtabular}
\caption{List of eigenvalues and eigenstates for single and two
excitation manifolds. Here $\chi_{\pm}=\sqrt{4g^2+(\Delta^2\pm2\mu)^2}$.}
\begin{tabular}{lccccccc}
 & Eigenvalues(shifts) &~~Eigenstates (Exciton polartions)\\
\hline
 &  & \\
Single excitation
&$\Delta/2+\mu$&$|1\rangle_{+}=[(\Delta+2\mu)|1,0\rangle+2ig|0,1\rangle]/\chi_{+}$\\
 manifold&  & \\
 &$\Delta/2-\mu$&$|1\rangle_{-}=\left[(\Delta+2\mu)|1,0\rangle+2ig|0,1\rangle\right]/\chi_{-}$\\
 &  & \\
 \hline
&  & \\
Two excitation
&$\Delta+2\mu$&$|2\rangle_{+}=[-i\sqrt{2}g|2,0\rangle+(\Delta+2\mu)|1,1\rangle+i\sqrt{2}g|0,2\rangle]/\chi_{+}$\\
manifold &  & \\
&$\Delta$&$|2\rangle_{0}=[-i\sqrt{2}g|2,0\rangle+\Delta|1,1\rangle+i\sqrt{2}g|0,2\rangle]/\mu$\\
  &  & \\
 &$\Delta-2\mu$&$|2\rangle_{-}=[-i\sqrt{2}g|2,0\rangle+(\Delta-2\mu)|1,1\rangle+i\sqrt{2}g|0,2\rangle]/\chi_{-}$\\
\end{tabular}
\end{ruledtabular}
\end{table*}

\begin{figure*}[t]
\includegraphics[width=16.5cm]{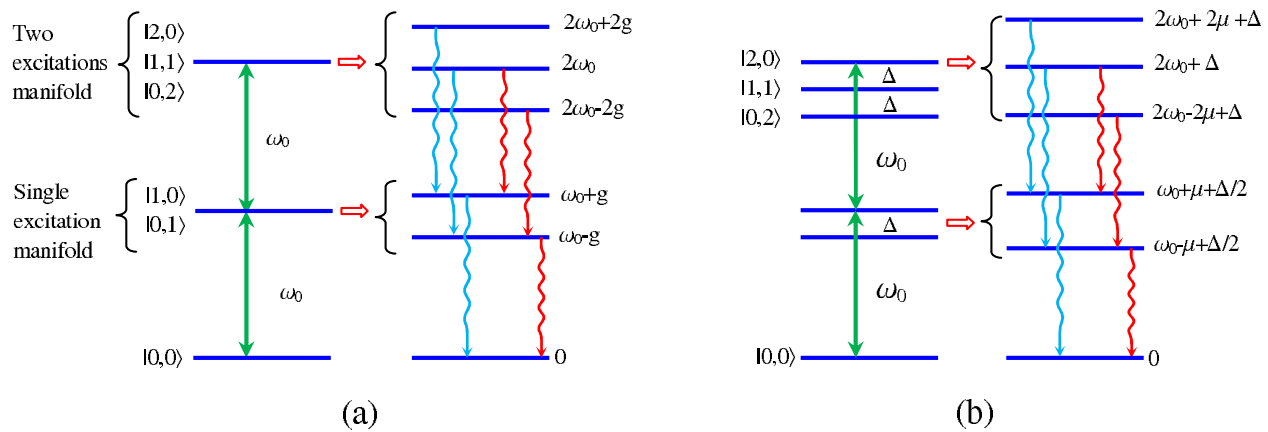}
\caption{(a) Dressed state energy level diagram for single and two
excitation manifolds when the exciton is at resonance with photon.
(b) Dressed states energy level diagram when the exciton frequency
is detuned by $\Delta$ from that of the photon. We have assumed
$\Delta$ to be positive for sake of simplicity. The bare states
$|n,m\rangle (n,m=0,1,2)$ represent $n$ numbers of excitons and $m$
numbers of photons. Even though there are 6 possible transitions
there are only 2 distinct transition frequencies namely:
$\omega-\omega_{0}=\mu+\Delta/2$ and
$\omega-\omega_{0}=-\mu+\Delta/2$, where
$\mu=\sqrt{g^2+\Delta^2/4}$.} \label{fig6}
\end{figure*}

In Fig. \ref{fig4} we plot the incoherent component of the power
spectrum as a function of scaled time $\gamma t$ for the cavity mode
initially in vacuum state and for the quantum well initially
containing one exciton. For zero detuning the power spectrum
consists of two well resolved peaks centered at
$\omega-\omega_{0}=\pm g$. This splitting can be understood from the
dressed state energy level diagram (see Fig \ref{fig6}(a)). Note
that for the case in which there is only one excitation, there are
two possible degenerate bare states: $|1,0\rangle$--one exciton and
no photon, and $|0,1\rangle$--one photon no exciton. However, the
strong exciton-photon coupling lifts the degeneracy of these two
bare states and results in two dressed states (exciton polaritons)
$|+\rangle=(|1,0\rangle+i|0,1\rangle)/\sqrt{2}$ and
$|-\rangle=(|1,0\rangle-i|0,1\rangle)/\sqrt{2}$ with eigenvalues $g$
and $-g$, respectively. In general, since the exciton-photon system
is coupled to the environment, exciton polartions are unstable
states. Thus the decay of the exciton and cavity photon leads to
exciton polaritons decay, which yields two peaks in the emission
spectrum.

 It is worth to note that even though we start off with a
single exciton in the quantum well, the cavity photons excite two or
more excitons in the quantum well. This results in more dressed
states in multi excitation manifolds. For example, as shown in Fig.
\ref{fig6}(a), for two excitation manifolds there are three dressed
states which are equally spaced in energy. This energy separation is
the same as the energy separation in one excitation manifold. Out of
the six possible transitions, from two excitations to single
excitation and then from single to ground state, there are only two
distinct frequencies. Therefore, the emission spectrum consists of
two peaks. This is different from the atom-photon coupling in which
the increase in excitation number increases the number of
emission spectrum peaks.

On the other hand, for nonzero detuning case, the emission spectrum
has two peaks whose centers are shifted to red (for positive
detuning). Here the one excitation bare states ($|1,0\rangle,
|0,1\rangle$) are separated by $\Delta$ and the two excitation
states ($|2,0\rangle, |1,1\rangle$, and $|0,2\rangle$) as well. The
eigenvalues and corresponding eigenstates are given in Table I. The
exciton-photon coupling leads to the generation of dressed states
(exciton polaritons). The decay of these states to the one
excitation and to the ground state gives rise to two emission peaks
whose frequencies are different from the zero detuning case as shown
in Fig. \ref{fig6} (b). Further, the density plot for the power
spectrum clearly shows that there are indeed two peaks separated by
$2\mu=2\sqrt{g^2+\Delta^2/4}$ as illustrated in Fig. \ref{fig5}.


\section{Second-order correlation function}

In this section we study the second-order correlation function of
the light emitted by the quantum well. Second-order correlation
function is a measure of the photon correlations between some time t
and a later time $t+\tau$. It is also an indicator of a quantum
feature that doesn't have classical analog. Quantum mechanically the
second-order correlation function is defined by
\begin{equation}\label{G1}
\text{g}^{(2)}\left( \tau \right) =\frac{\left\langle b^{\dagger}
(t)b^{\dagger} (t+\tau )b(t+\tau )b(t)\right\rangle_{ss}
}{\left\langle b^{\dagger} (t)b(t)\right\rangle_{ss} ^{2}}.
\end{equation}
The correlation function that appears in \eqref{G1} can be obtained
using the solution \eqref{m2} and together with the properties of
the Langevin noise forces. Note that as the mean values of the
noise forces are zero and the Langevin equations are linear we apply
the Gaussian properties of the noise forces. To this end, using Eq.
\eqref{m2}, we obtain

\begin{figure}[t]
\includegraphics{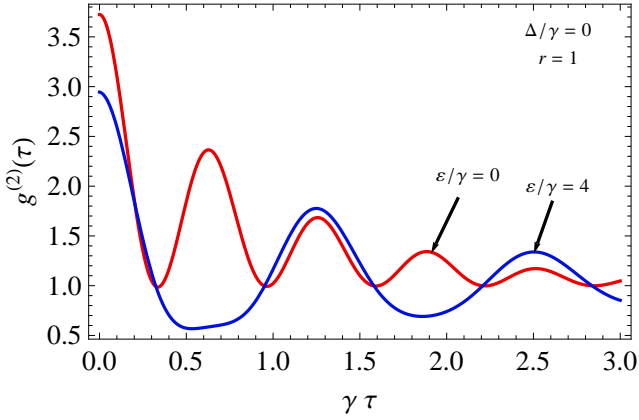}
\caption{Plot of second-order correlation function versus normalized
time $\gamma \tau$ for $g/\gamma=5$, $\kappa/\gamma=1.2$,
$\Delta/\gamma=0$, squeezing parameter $r=1$, and for different
value of pump amplitude $\varepsilon/\gamma$.}\label{g22}
\end{figure}
\begin{align}\label{g1}
    \text{g}^{(2)}(\tau)=1&+\frac{1}{\langle
    b^{\dagger}b\rangle_{ss}^2}\Big[ \frac{\kappa^2}{4}\left(4M^2 A_{3}+N^2
A_{2}^2\right)e^{-2\Gamma\tau}\notag\\
    &+\frac{\kappa
    \varepsilon^2}{g^2}\left[M A_{1}+N A _{2}\cos(\Delta
    \tau/2)\right]e^{-\Gamma\tau}\Big],
  \end{align}
where
\begin{align}\label{g2}
   A_{1}(\tau)&=\frac{2\mu\cos\mu\tau\sin(\Delta \tau/2)-\Delta\cos(\Delta
    \tau/2)\sin\mu\tau}{4\mu^2}\notag\\
    &+\frac{g^2\cos\mu \tau[2\Gamma\cos(\Delta t/2)-\Delta\sin(\Delta
    t/2)]}{\mu^2[\Delta^2+4\Gamma^2]},
\end{align}
\begin{equation}\label{g3}
    A_{2}(\tau)=\frac{g^2}{2\Gamma\mu^3}\left(\mu
    \cos\mu\tau+\Gamma\sin\mu\tau\right),
\end{equation}
\begin{equation}\label{g4}
    A_{3}(\tau)=\frac{4\mu^2\cos^2\mu\tau+\Delta^2\sin^2\mu\tau+4\Gamma\mu\sin(2\mu\tau)}{16\mu^2[\Delta^2+4\Gamma^2]},
\end{equation}
and $\langle b^{\dagger}b\rangle_{ss}$ is given by \eqref{m2}. The
term in Eq. \eqref{g1} represents the second order correlation
function for the coherent light. This can easily be seen by setting
$N=M=0$. The first term in the square bracket is the contribution to
the second-order correlation function from the squeezed vacuum
reservoir while the second term describes the interference term
between the coherent field and the reservoir. Note that at
$\tau\rightarrow \infty$ $\text{g}^{(2)}$ becomes unity as it should
be, showing no correlation between the photons.

The dynamical behavior of the second-order correlation function is
illustrated in Fig. \ref{g22}. We see from this figure that the
correlation function shows oscillatory behavior with oscillation
frequency equal to the photon-exciton coupling constant (g) for zero
detuning case and in the absence of the external driving field.
However, the frequency of oscillation is reduced by a factor of 1/2
in the presence of external coherent field.

It easy to see that $\text{g}^{(2)}(0)$ is always greater than unity
indicating photon bunching. This is in contrary to what has been
observed in atomic cavity QED, where the photons show antibunching
property \cite{Set08}. This is due to the fact that there is a
finite time delay between absorbtion and subsequent emission of a
photon by the atom. In the case of semiconductor cavity QED,
however, the cavity photons can excite two or more excitons at the
same time depending on the number of photons in the cavity leading
to possible multi photon emission. This is the reason why the
photons emitted by excitons are bunched. Indeed, excitation of two
or more excitons in the quantum well is shown in Figs.
\ref{fig1}-\ref{fig3}.

\section{Quadrature squeezing}
Next we study the squeezing properties of the fluorescent light by
evaluating the variances of the quadrature operators. The variances
of the quadrature operators for the fluorescent light is given by
\begin{equation}\label{Q1}
    \Delta b_{\pm}^2=1+2\langle b^{\dagger}b\rangle\pm\left[\langle
    b^{2}\rangle+\langle b^{\dagger 2}\rangle\right]\mp(\langle
b^{\dagger}\rangle\pm\langle b\rangle)^2,
\end{equation}
where $b_{+}= (b^{\dagger}+b)$ and $b_{-}=i(b^{\dagger}-b)$. It can
be easily seen from this definition that the quadrature operators
satisfy the commutation relation $[b_{+}, b_{-}] = 2i$. It is well
known that for the fluorescent light to be squeezed the variances of
the quadrature operators should satisfy the condition that either
$\Delta b_{+}^2 < 1$ or $\Delta b_{-}^2 < 1$. Using Eq. \eqref{a2}
and the properties of the noise operators we find the variances to
be
\begin{align}\label{Q3}
    \Delta b_{\pm}^2&=1+\frac{\kappa N
    g^2}{2\Gamma\mu^2}\pm\frac{2\kappa M
    g^2\Gamma}{\mu^2(\Delta^2+4\Gamma^2)}\notag\\
    &+\Big[2\lambda_{1}(t)+2\kappa N\lambda_{2}(t)
    \pm \kappa
    M\lambda_{4}\Big] e^{-2\Gamma t},
    \end{align}
in which $\lambda_{1}$ and $\lambda_{2}$ are same as defined earlier
in section (III) and
\begin{align*}
    \lambda_{4}&=\frac{g^2}{\mu^2}\Bigg[\frac{\Delta\sin(\Delta
    t/2)-2\Gamma\cos(\Delta t)}{\mu^2(\Delta^2+4\Gamma^2)}\notag\\
    &-\frac{\sin[(\Delta-2\mu)t]}{2((\Delta-2\mu))}-\frac{\sin[(\Delta+2\mu)t]}{2((\Delta+2\mu))}\Bigg]
\end{align*}
It is straightforward to see that in the steady state the variances
reduce to
\begin{align}\label{Q5}
\Delta b_{+}^2&=1+\frac{\kappa N
    g^2}{2\Gamma\mu^2}+\frac{2\kappa M
    g^2\Gamma}{\mu^2(\Delta^2+4\Gamma^2)},
\end{align}
\begin{align}\label{Q6}
\Delta b_{-}^2&=1+\frac{\kappa N
    g^2}{2\Gamma\mu^2}-\frac{2\kappa M
    g^2\Gamma}{\mu^2(\Delta^2+4\Gamma^2)}.
\end{align}

\begin{figure}[t]
\includegraphics{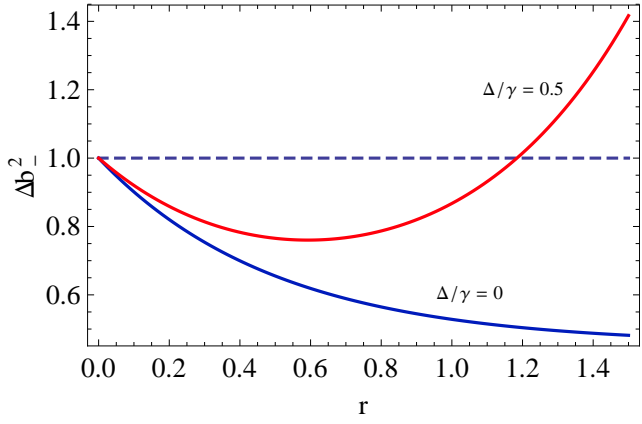}
\caption{Plots of the steady state quadrature variance [\eqref{Q6}]
vs squeeze parameter $r$ for $g/\gamma=5$, $\kappa/\gamma=1.2$, and
for the different values of exciton-photon detuning $\Delta/\gamma$
.}\label{qv2}
\end{figure}
\begin{figure}[t]
\includegraphics{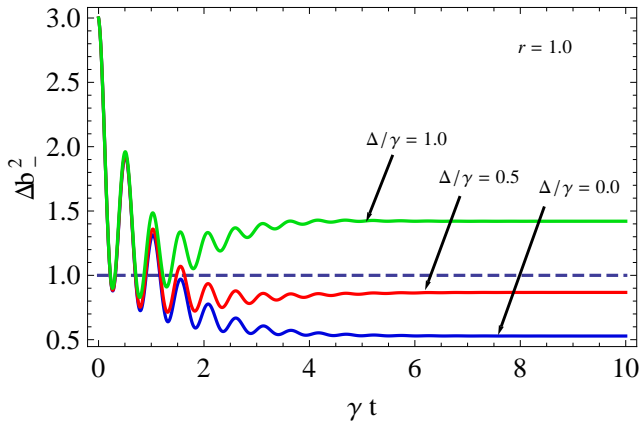}
\caption{Plots of the quadrature variance [\eqref{Q3}] vs scaled
time $\gamma t$ for $g/\gamma=5$, $\kappa/\gamma=1.2$, squeeze
parameter $r=1$ and for the different values of exciton-photon
detuning $\Delta/\gamma$.}\label{qv1}
\end{figure}

From the above expressions we find that, in the steady state, the
quadrature variances crucially depend on the detuning, the
cavity-exciton coupling strength and amount of squeezing provided by
the reservoir. Further, it is apparent that if there is any
squeezing it can only be present in the $b_{-}$ quadrature. Thus for
rest of this section we will only discuss the properties of variance
in the $b_{-}$ quadrature. As a special case, we consider that the
cavity mode to be at resonance with with the excitonic transition
frequency and put $\Delta = 0$ in Eq. (\ref{Q6}). We then find that
\begin{align}\label{Q7}
 \Delta b_{-}^{2} = 1-\frac{\kappa}{\kappa+\gamma}(1-e^{-2r}) < 1.
 \end{align}
Equation (\ref{Q7}) then suggest that higher squeezing in the
reservoir leads to better squeezing  of the fluorescent light. In
Fig. \ref{qv2} we confirm this behavior by plotting the steady state
$\Delta b_{-}^2$ as a function of the squeezing parameter $r$.
In addition, from Eq. (\ref{Q7}) we see that as $e^{-2r} \rightarrow 0$
quickly with increase in $r$ the maximum possible squeezing
achievable in our system is $ 50\%$ for $\kappa = \gamma$. This is
also depicted to be true in Fig. \ref{qv2}.

In presence of detuning $(\Delta \neq 0)$ the behavior of $\Delta
b_{-}^2$ changes dramatically. We find that for some small detuning
$\Delta/\gamma = 0.5$ there exist a range of the squeezing parameter
$r~(0 < r < 1.3)$ where one can see squeezing of the fluorescent
light emitted from the exciton, however for higher values of $r$ it
vanishes. This thus implies that in presence of detuning stronger
squeezing of the reservoir leads to negative effect on the squeezing
of the emitted radiation from the excitons. In Fig. \ref{qv1}, we
plot the time evolution of the quadrature variance $\Delta b_{-}^2$
(Eq. \ref{Q3}) as a function of the normalized time $\gamma t$ for
$r = 1$ and different values of detuning. It is seen, in general,
that the variance oscillates initially with the amplitude of
oscillation gradually damping out at longer time. Eventually, at
large enough time the variance becomes flat and approaches to the
steady state value. Interestingly, our results show that even though
there is no squeezing of the fluorescent light at the initial
moment, for small or zero detuning, transient squeezing gradually
develops.  Moreover, we also find that for weak squeezing of the
reservoir, even in presence of small detuning, the initial transient
squeezing is sustained and finally leads to a steady state
squeezing. This can be understood as a consequence of strong
interaction of the quantum well with the squeezed photon entering
via the cavity mirror. In case of large detuning the exciton is
unable to absorb photons from the squeezed reservoir and thus no
squeezing develops in the fluorescence.

\section{Conclusion}
In this paper we consider a semiconductor quantum well in a cavity
driven by external coherent light and coupled to a single mode
squeezed vacuum reservoir. We study the photon statistics and
nonclassical properties of the light emitted by the quantum well in
the presence of exciton-photon detuning in the strong coupling
regime. The effects of coherent light and the squeezed vacuum
reservoir on the intensity of the fluorescence are quite different.
The former leads to a transient peak intensity which eventually
decreases to a considerably smaller steady state value. In contrast,
the latter, however, gives rise to a gradual increase in the
intensity and leads to maximum intensity at steady state. This
difference is attributed to the nature of photons that the two
sources produce. As a signature of strong coupling between the
excitons in the quantum well and cavity photons the emission
spectrum consists of two peaks corresponding to the two
eigenenergies of the dressed states. Further, we find that the
fluorescence exhibit nonclassical feature--quadrature squeezing--as
a result of strong interaction of the excitons with the squeezed
photons entering via the cavity mirror. In view of recent successful
experiments on Autler-Townes effect in GaAs/AlGaAs \cite{Wag2010}
and gain without inversion in semiconductor nanostructures
\cite{Fro2006}, the quantum statistical properties of the
fluorescence emitted by the quantum well can be tested
experimentally.
\begin{acknowledgements}
One of the authors (E.A.S.) is supported by a fellowship from Herman
F. Heep and Minnie Belle Heep Texas A$\&$M University Endowed Fund
held/administered by the Texas A$\&$M Foundation.
\end{acknowledgements}

\end{document}